\documentclass[preprint2]{aastex}

\shorttitle{Distribution of PAH emission in NGC~4589}
\shortauthors{Kaneda et al.}

\begin{document}


\title{PAH emission from the dust lane of an elliptical galaxy with the Spitzer IRS}


\author{Hidehiro Kaneda\altaffilmark{1}, Takashi Onaka\altaffilmark{2}, Itsuki Sakon\altaffilmark{2}, Tetsu Kitayama\altaffilmark{3}, Yoko Okada\altaffilmark{4}, Toyoaki Suzuki\altaffilmark{5}, Daisuke Ishihara\altaffilmark{1}, Mitsuyoshi Yamagishi\altaffilmark{1}}
\altaffiltext{1}{Graduate School of Science, Nagoya University, Chikusa-ku, Nagoya 464-8602, Japan}
\email{kaneda@u.phys.nagoya-u.ac.jp}
\altaffiltext{2}{Graduate School of Science, The University of Tokyo, Bunkyo-ku, Tokyo 113-0033, Japan}
\altaffiltext{3}{Department of Physics, Toho University, Funabashi, Chiba 274-8510, Japan}
\altaffiltext{4}{I. Physikalisches Institut, Universit\"at zu K\"oln, 50937 K\"oln, Germany}
\altaffiltext{5}{Institute of Space and Astronautical Science, Japan Aerospace Exploration Agency, Sagamihara, Kanagawa 252-5210, Japan}


\begin{abstract}
{\it Spitzer} and {\it AKARI} observations have found that polycyclic aromatic hydrocarbons (PAHs) are present in nearby elliptical galaxies, but their spatial distributions are still unknown. In order to investigate their distributions, we performed deep spectral mapping observations of the PAH-detected elliptical galaxy NGC~4589, a merger remnant with a minor-axis optical dust lane. As a result, we obtain clear evidence that the PAH 11.3 $\mu$m emission comes predominantly from the dust lane of the galaxy. We also detect molecular hydrogen line emissions from the dust lane. The PAH 17 $\mu$m emission is distributed differently from the PAH 11.3 $\mu$m emission, and more similarly to the dust continuum emission. From their distinctive distributions, we suggest that the PAHs responsible for the 11.3 $\mu$m feature are secondary products through the evolution of the ISM brought in by the merger.    
\end{abstract}
\keywords{ISM: lines and bands --- dust, extinction --- galaxies: elliptical and lenticular, cD --- galaxies: individual(NGC~4589) --- infrared: galaxies}

\section{Introduction}
Many elliptical galaxies are not ISM-free, which contain a considerable amount of cool dust (e.g. Knapp et al. 1989; Goudfrooij \& de Jong 1995; Temi et al. 2004; Temi et al. 2007). Early {\it Spitzer} IRS observations have found that even polycyclic aromatic hydrocarbons (PAHs) are present in elliptical galaxies (Kaneda et al. 2005). The spectra show that the PAH 11.3 $\mu$m feature is notably strong relative to usually the strongest 7.7 $\mu$m feature, which might be typical in evolved systems such as elliptical galaxies, i.e. the dominance of neutral PAHs over ionized ones due to very soft radiation fields from evolved stars. In addition to the PAH features, a series of H$_{2}$ rotational lines as well as ionic fine-structure lines (e.g. [NeII] and [NeIII]) are detected from a large fraction of the elliptical galaxies, implying the presence of warm molecular and ionized gases \citep{Kan08a}. Their presence seems to be inconsistent with the dominance of neutral PAHs irradiated by soft radiation field. 

What are the origin and fate of the PAHs in elliptical galaxies? How are they related with the warm molecular and ionized gases as well as the cool dust? How do their properties change from the center to outer regions? Spatial information is indispensable to disentangle various competing components and tackle these issues. To pursue follow-on studies of the early {\it Spitzer} discoveries, we performed deep spectral mapping observations of elliptical galaxies with the {\it Spitzer} IRS.  In this Letter, we report the result of NGC~4589, a nearby E2 elliptical galaxy with a minor-axis optical dust lane \citep{Mol89}. NGC~4589 has a complex stellar rotation field, although its morphology shows the smooth optical profile following the de Vaucouleurs law \citep{Mol89}, which supports that the galaxy is a relatively old merger remnant. Recently, a core-collapse supernova (SN Ib 2005cz) was observed in this genuine elliptical galaxy \citep{Hak08}; SN Ib 2005cz is likely to be the end product of one of young stars, which were produced by the galaxy merger about $10^8$ years ago \citep{Kaw10}. NGC~4589 is an X-ray-emitting elliptical galaxy with a low-ionization nuclear emission-line region (LINER) nucleus \citep{OSu01, Dre85, Wil85}. 

Our earlier IRS observation detected the PAH features from the galaxy in a staring mode but not in a mapping mode, and therefore we could not spatially resolve them \citep{Kan08a}. After the {\it Spitzer} detection of the PAH features, their spatial distributions were estimated from the {\it AKARI} 11 $\mu$m and 15 $\mu$m band images, while that of the far-infrared (IR) dust was obtained from the {\it AKARI} 90 $\mu$m image \citep{Kan08b}. They all show significant deviations from the smooth stellar distribution, from which Kaneda et al. (2008b) suggested that the PAHs and far-IR dust are mostly of an interstellar origin, rather than of a stellar origin.
In this Letter, from the {\it Spitzer}/IRS spectral mapping of NGC~4589, we clearly find that the PAH 11.3 $\mu$m emission comes from the minor-axis dust lane, while the PAH 17 $\mu$m emission is distributed differently from the PAH 11.3 $\mu$m emission, and more similarly to the far-IR dust emission. 

\section{Observations}
We observed \dataset[ADS/Sa.Spitzer#26091008]{NGC~4589} in a spectral mapping mode with the IRS (Houck et al. 2004) onboard {\it Spitzer} \citep{Wer04} on May 6--7, 2009. The total observation time was 7.7 hours. The data were taken in part of our Guest Observers (GO5) program (PI: H. K.; program ID: 50369). Starting with the basic calibrated data (BCD; pipeline version S18.7.0) reduced by the {\it Spitzer} Science Center (SSC), we extracted spectral image data by the CUBISM software (version 1.7) following standard procedures.

Figure 1a shows the slit positions of the Short-Low (SL; 5.2--14.5 $\mu$m) and Long-Low (LL; 14--36 $\mu$m) modules. We made a $33\times 2$ grid for SL with a ramp duration of 60 s for each exposure, and a $13\times 2$ grid for LL with a ramp duration of 30 s for each. We shifted the slit in directions perpendicular and parallel to its length by $1''.85$ and $3''.0$, respectively, for SL, and by $5''.1$ and $9''.0$, respectively, for LL. The 50 \% overlap of the slit aperture in the perpendicular direction is crucial for reliable estimation of the slit-loss correction factors for extended sources, which we could not obtain in our earlier IRS staring observations. The SL and LL slit directions are almost perpendicular to each other, resulting in the coverage of $\sim 1'\times 1'$ area around the center of the galaxy by both modules. We repeated the same spectral mapping observations two and three times for SL and LL, respectively. 

\begin{figure}[h]
\epsscale{1.07}
\plottwo{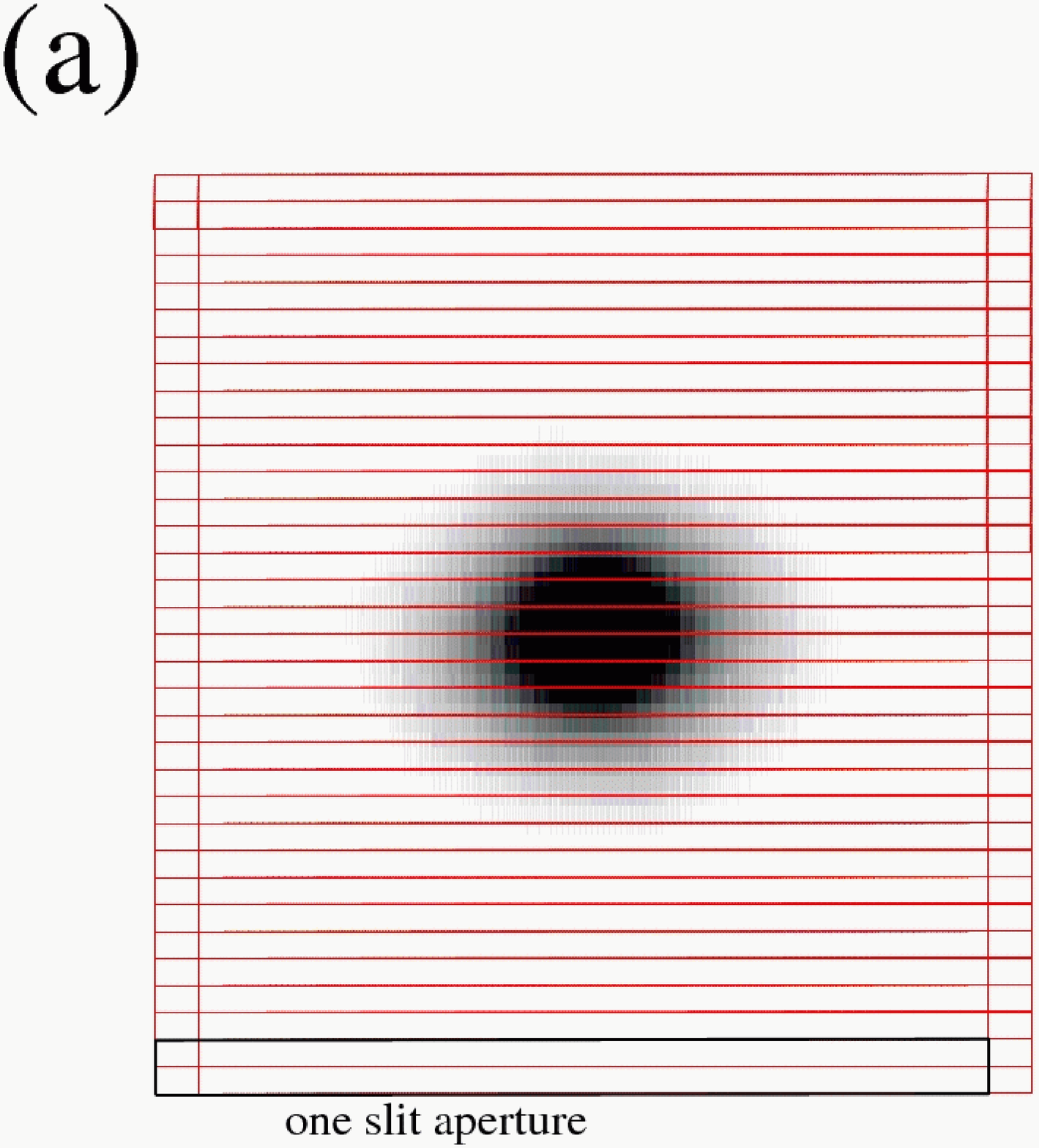}{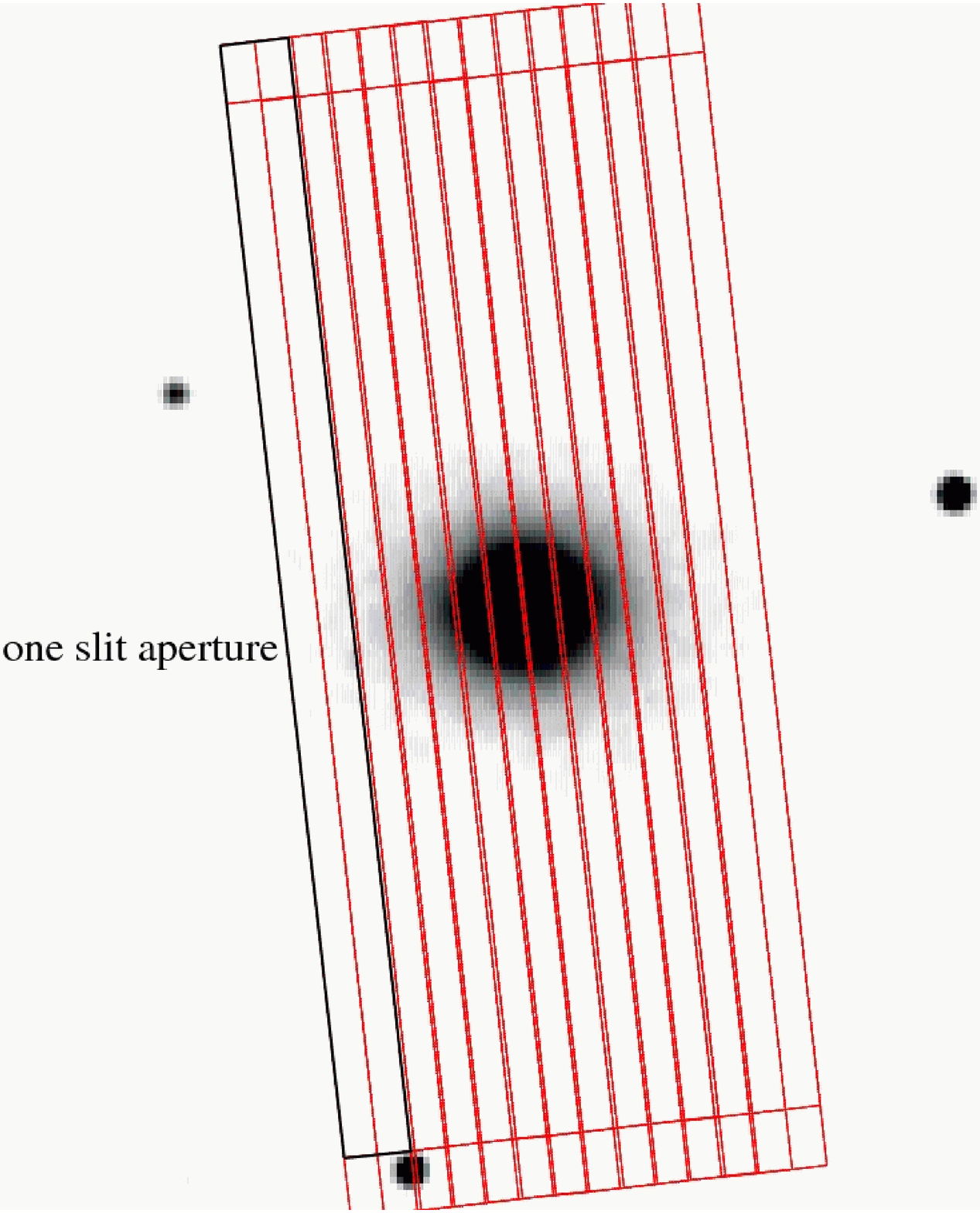}
\epsscale{1.0}
\plotone{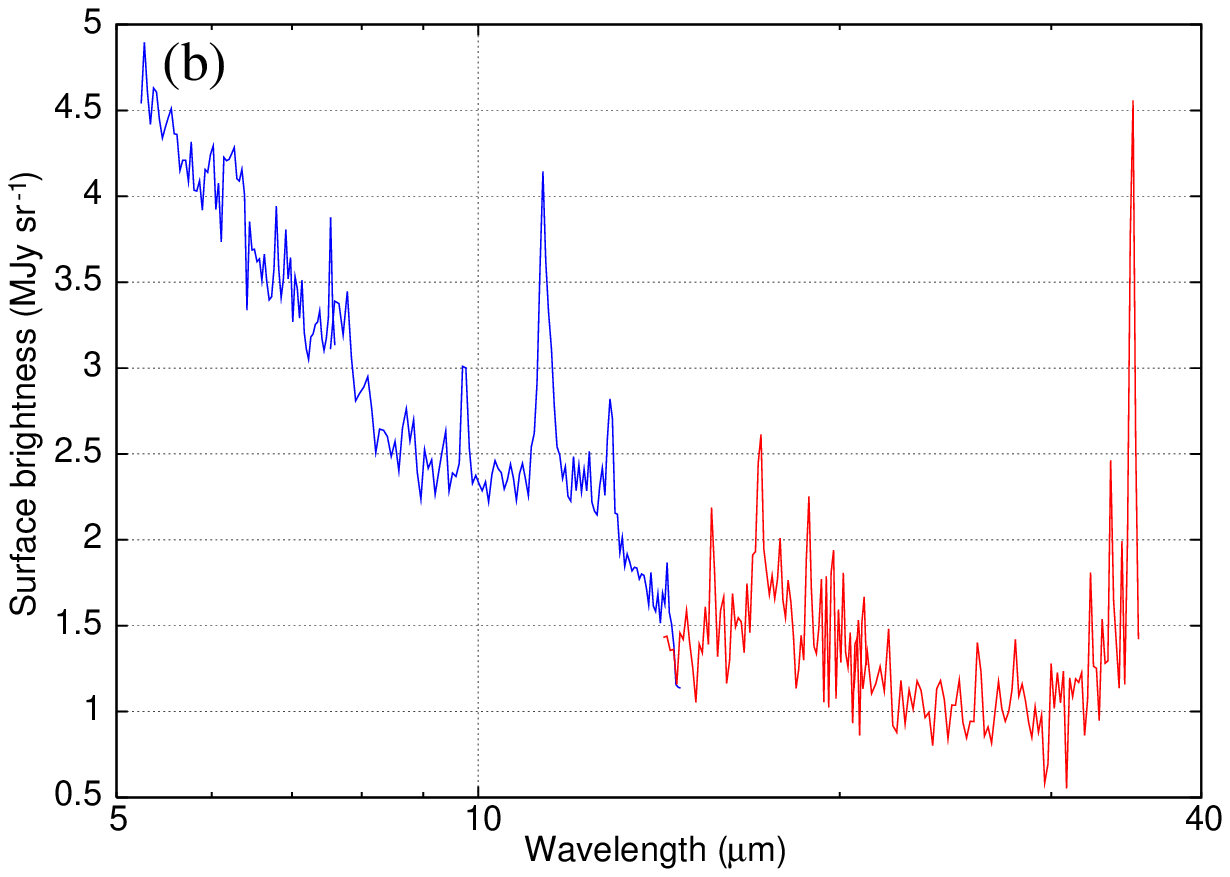}
\caption{(a) illustrative views of IRS SL (left) and LL (right) slit positions for the spectral mapping observations of NGC~4589, superposed on the {\it 2MASS} J-band image of NGC~4589. The image sizes are $1'.5\times 1'.5$ for SL and $3'\times 3'$ for LL. The north is up and the east is to the left. (b) The background-subtracted IRS SL (blue) and LL (red) spectra of the central $15''$ region of NGC~4589.}
\end{figure}

From the spectral image data, we created the spectra of the central $15''$ region. Figure 1b shows the resultant spectra, where the background spectra were created from eight slit-aperture data at the ends of the spectral mapping area for SL and four slit-aperture data for LL, and subtracted.
In Fig.1b, we clearly see the prominent PAH 11.3 $\mu$m feature and the PAH 17 $\mu$m broad feature, together with several emission lines such as H$_2$S(3) at 9.7 $\mu$m, [NeII] at 12.8 $\mu$m, [NeIII] at 15.6 $\mu$m, H$_2$S(1) at 17.0 $\mu$m, [SiII] at 34.8 $\mu$m, and the mid-IR dust continuum emission at wavelengths longer than 30 $\mu$m. Photospheric emission from old stars appears to dominate in the continuum emission at wavelengths shorter than 20 $\mu$m, judging from the slope of the continuum. Hereafter the bin sizes of spectral images are set to be $1''.85$ for SL and $5''.08$ for LL. We applied smoothing with a boxcar kernel of 3 pixels in width ($\sim 5''$ for SL and $\sim 15''$ for LL) for every spectral map. In table 1, we list the wavelength ranges used for extracting each spectral component, together with the flux from the central $45''$ region. The [NeII] and H$_2$S(1) line emissions are blended with the underlying PAH 12.7 $\mu$m and 17 $\mu$m features, respectively.

\begin{deluxetable}{lrrrr}
\tabletypesize{\scriptsize}
\tablecaption{Wavelength ranges and fluxes of spectral components}
\tablewidth{0pt}
\tablehead{
Component & Signal & Background&Flux\tablenotemark{a}&Notes\\
 & ($\mu$m) & ($\mu$m)&
}
\startdata
$5.5-6.5$ $\mu$m continuum&$5.5-6.5$&\dots&47&  \\
H$_2$S(3) &$9.6-9.8$&$9.2-9.6$, $9.9-10.5$&2.6& \\
PAH 11.3 $\mu$m&$11.0-11.8$&$9.9-11.0$, $11.8-12.5$&11&  \\
PAH 12.7 $\mu$m&$12.6-12.9$&$13.0-13.2$&2.5&  \\
$[$NeII$]$ &$12.9-13.0$&$13.0-13.2$&1.7&blended with PAH 12.7 $\mu$m \\
$[$NeIII$]$ &$15.6-15.8$&$14.4-15.5$, $15.8-16.6$&0.7& \\
H$_2$S(1) &$17.0-17.3$&$14.4-15.5$, $15.8-16.6$, $19.0-20.9$&2.4&blended with PAH 17 $\mu$m\\
PAH 17 $\mu$m&$17.3-18.7$&$14.4-15.5$, $15.8-16.6$, $19.0-20.9$&2.9&\\
$30-35$ $\mu$m continuum&$30.0-34.9$&\dots&18&\\
$[$SiII$]$ &$34.9-35.3$&$30.0-34.9$&3.0& \\
\enddata
\tablenotetext{a}{The fluxes from the central region of $45''$ in diameter. The values are given in units of $10^{-3}$ Jy for the continuum emissions and $10^{-17}$ W m$^{-2}$ for the feature and the line emissions.}

\end{deluxetable}

\section{Results}
Figure 2 shows the contour maps of the PAH 11.3 $\mu$m and 17 $\mu$m emission features overlaid on the contour maps of the $5.5-6.5$ $\mu$m and $30-35$ $\mu$m continuum emissions, respectively. The $5.5-6.5$ $\mu$m continuum emission shows a smooth stellar distribution, which is consistent with the {\it 2MASS} image in Fig.1a. However, the PAH 11.3 $\mu$m emission exhibits a distinctly elongated distribution, posessing an excellent spatial coincidence with the minor-axis optical dust lane (see Fig.2 of M\"ollenhoff \& Bender 1989). Thus we for the first time obtain direct evidence that the PAH emission comes predominantly from the optical dust lane of the elliptical galaxy. 

The PAH 17 $\mu$m feature shows a spatial distribution different from that of the PAH 11.3 $\mu$m emission, apparently extending to the southeast and the west. Since the PAH 17 $\mu$m emission has a relatively small feature-to-continuum ratio, the distribution may be sensitive to uncertainties in the background subtraction. As plotted in Fig.2b, however, the $14-21$ $\mu$m background continuum shows a smooth elliptical distribution entirely different from the PAH 17 $\mu$m feature and similar to the $5.5-6.5$ $\mu$m continuum, supporting that the stellar emission dominates in the continuum at $<$ 20 $\mu$m. Thus the above result is not affected by the uncertainties in the background subtraction. Both PAH distributions in Fig.2 are consistent with the earlier estimates based on the {\it AKARI} multi-band images (see Fig.3 of Kaneda et al. 2008b). The $30-35$ $\mu$m continuum emitted by very small dust grains also shows a hint of extension to the southeast, which is, at least, not of an elliptical shape as the stellar continuum emissions show. 

The 11.3 $\mu$m feature is attributed to a C-H out-of-plane bending mode mainly arising from neutral PAHs \citep{All89, Dra07, Tie08}. The 17 $\mu$m broad feature is interpreted as the emission of larger PAHs with a few thousand carbon atoms in  C--C--C in- and out-of-plane bending modes \citep{Van00, Pee04, Tie08}. In general, larger PAHs can be excited by softer radiation field, which might cause the observed different distributions. However such soft radiation as optical emission from old stars is hard enough to excite PAHs with the number of C atoms of $100-1000$ \citep{Sch93}, which are main emitters of the 11.3 $\mu$m feature. The different distributions are therefore difficult to be explained only by variations in the radiation hardness. It is likely that the PAHs responsible for the 17 $\mu$m feature belong to a different population from those for the 11.3 $\mu$m feature. 

\begin{figure}
\epsscale{.95}
\plotone{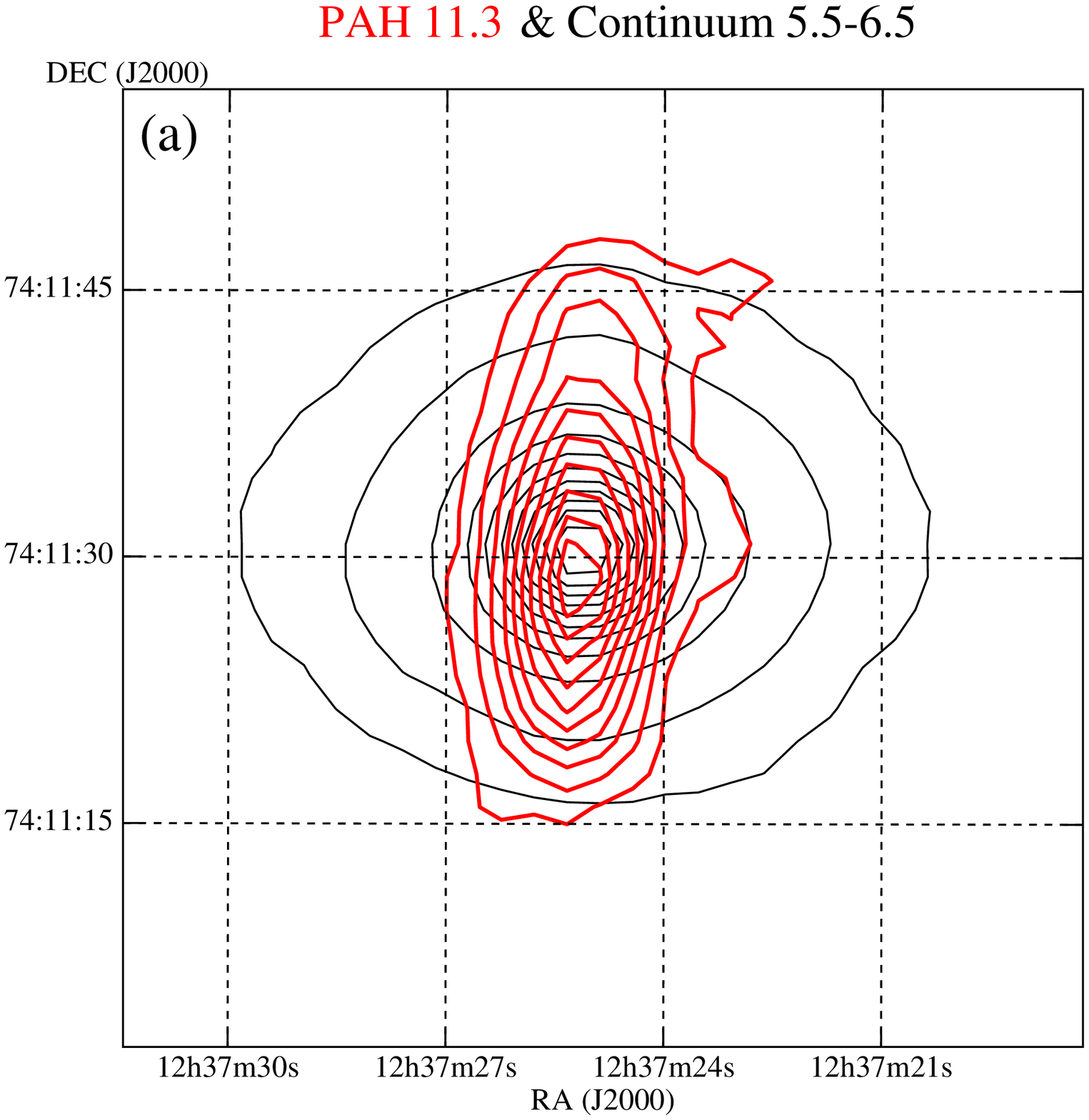}\\
\plotone{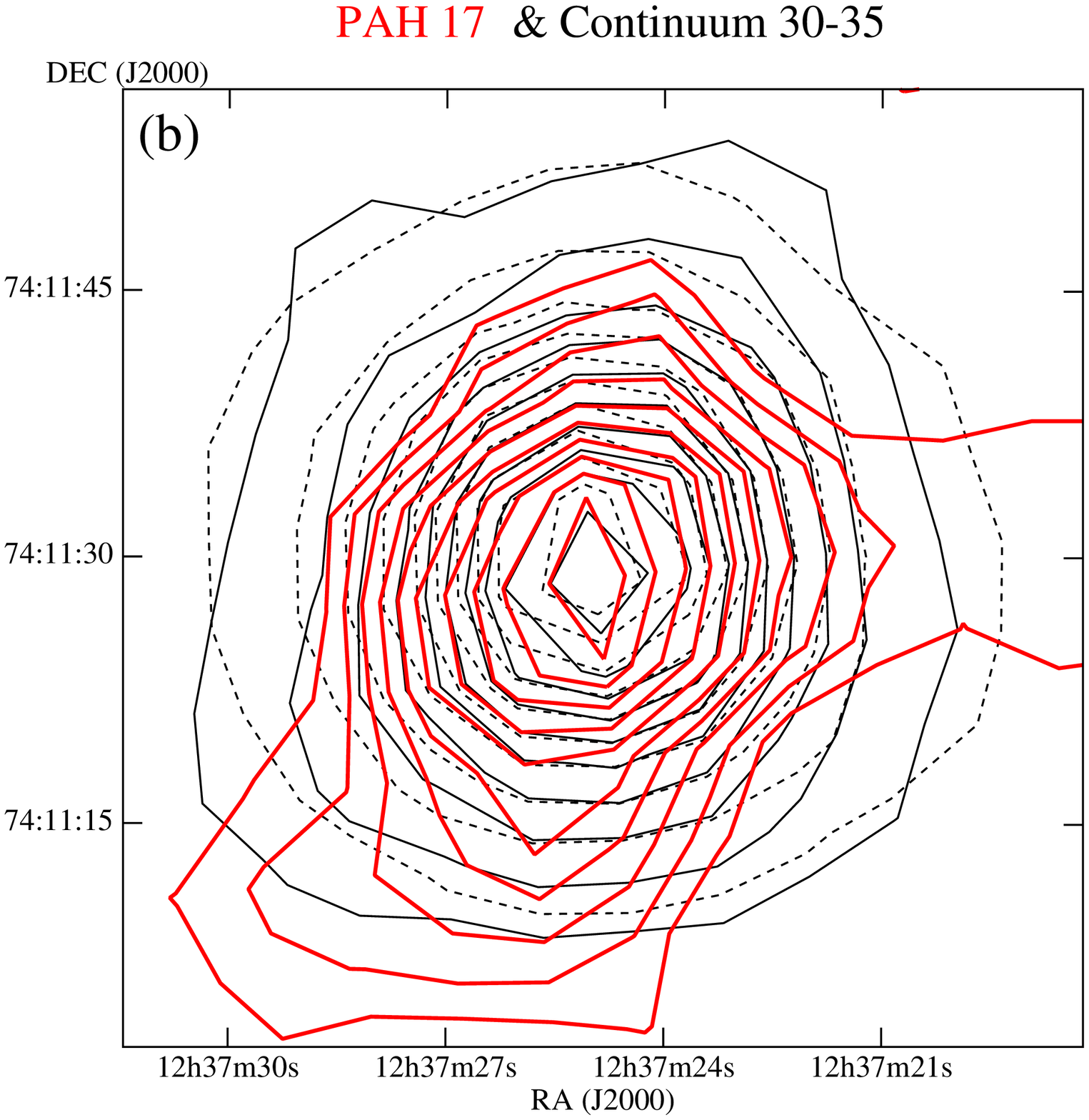}
\caption{(a) contour map of the PAH 11.3 $\mu$m feature (thick red lines) on that of the $5.5-6.5$ $\mu$m continuum emission (thin black lines). (b) The contour map of the PAH 17 $\mu$m feature (thick red lines) on that of the $30-35$ $\mu$m continuum (thin black lines). The contours of the $14-21$ $\mu$m background continuum are shown together (thin dashed lines). For each map, the contours are drawn on a linear scale from 10 to 90 \% of the peak ($5.5-6.5$ $\mu$m continuum: 11.6 MJy sr$^{-1}$, PAH 11.3 $\mu$m: $3.1\times 10^{-8}$ W m$^{-2}$ sr$^{-1}$, PAH 17 $\mu$m: $2.8\times 10^{-9}$ W m$^{-2}$ sr$^{-1}$, $14-21$ $\mu$m continuum: 1.4 MJy sr$^{-1}$, $30-35$ $\mu$m continuum: 1.5 MJy sr$^{-1}$). Only for the $5.5-6.5$ $\mu$m continuum map is the 5 \% level contour added. The lowest contours correspond to levels with S/Ns of 12 ($5.5-6.5$ $\mu$m continuum), 5.5 (PAH 11.3 $\mu$m), 2.6 (PAH 17 $\mu$m), 4.5 ($14-21$ $\mu$m continuum), and 4.7 ($30-35$ $\mu$m continuum), where the noises are estimated from nearby blank-sky fluctuations.}
\end{figure}

\begin{figure}
\epsscale{1.07}
\plottwo{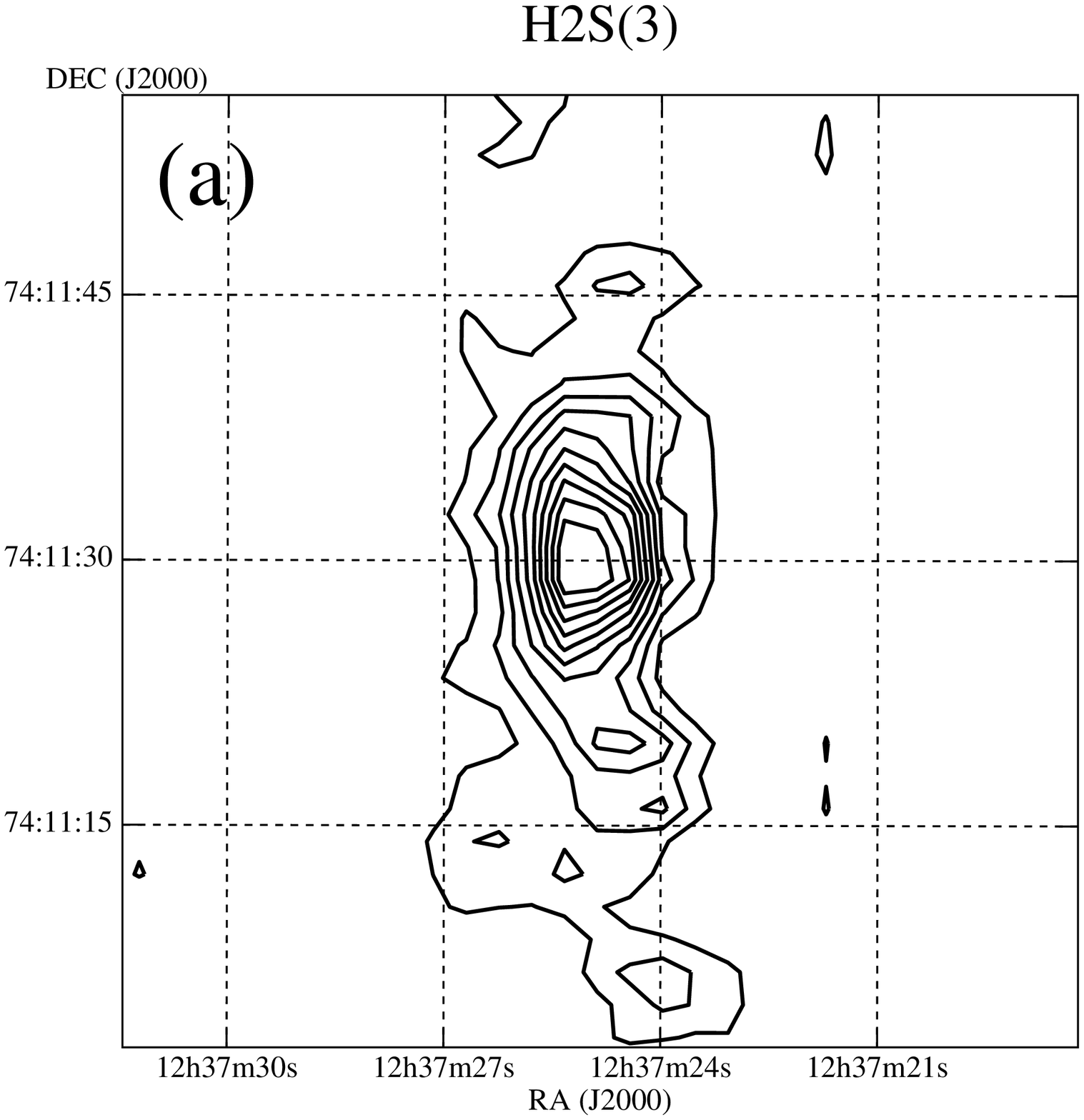}{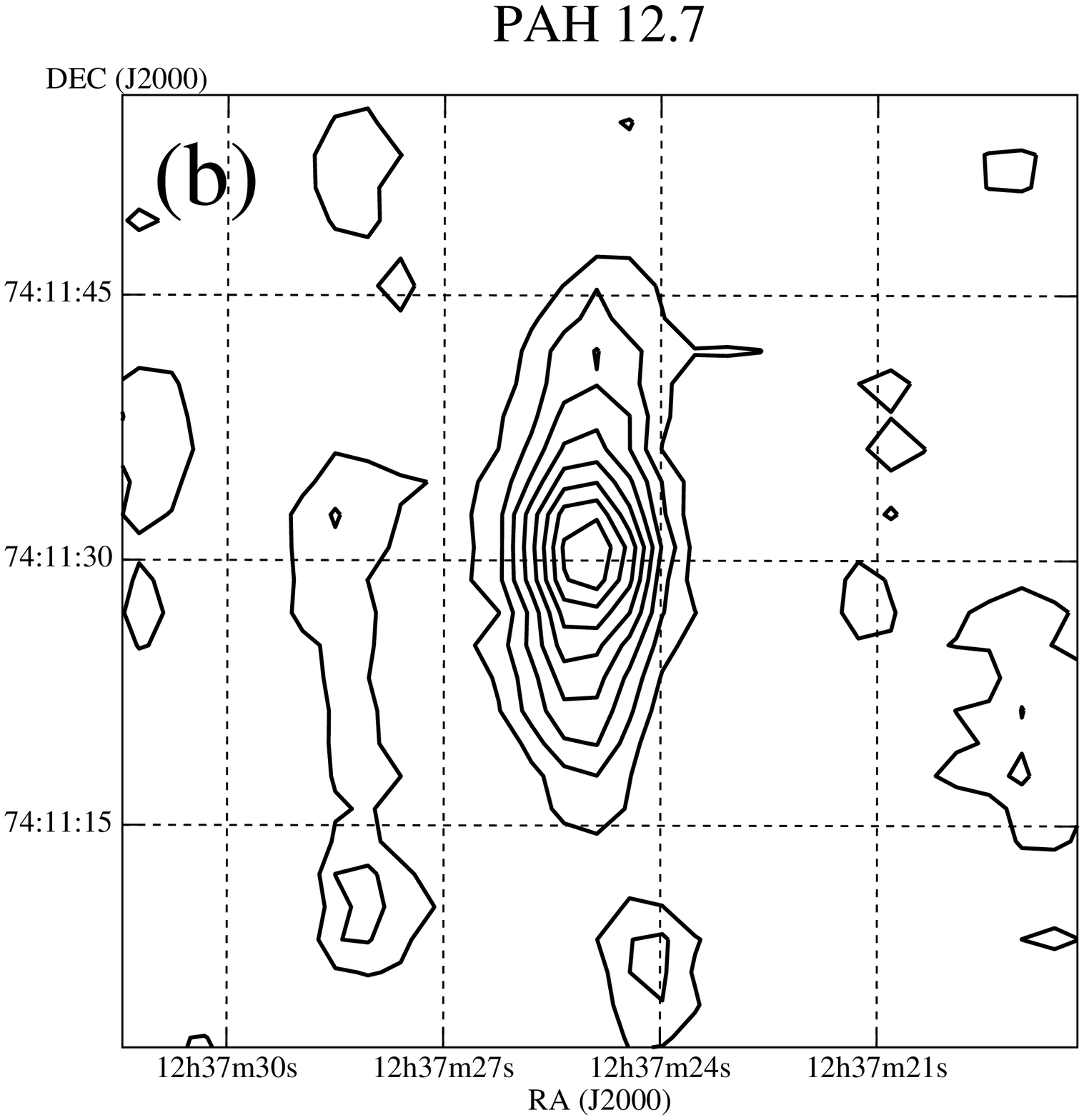}

\plottwo{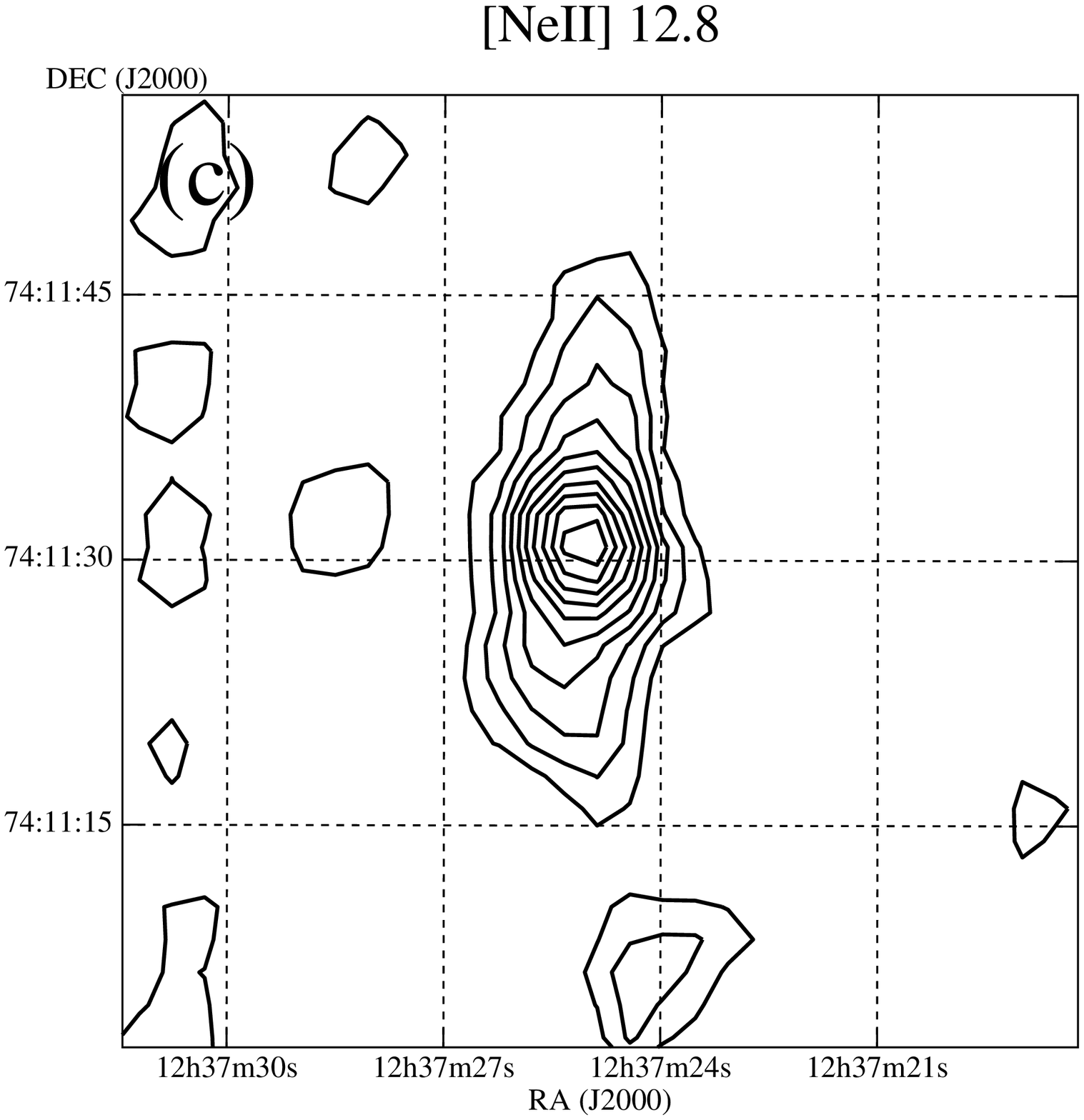}{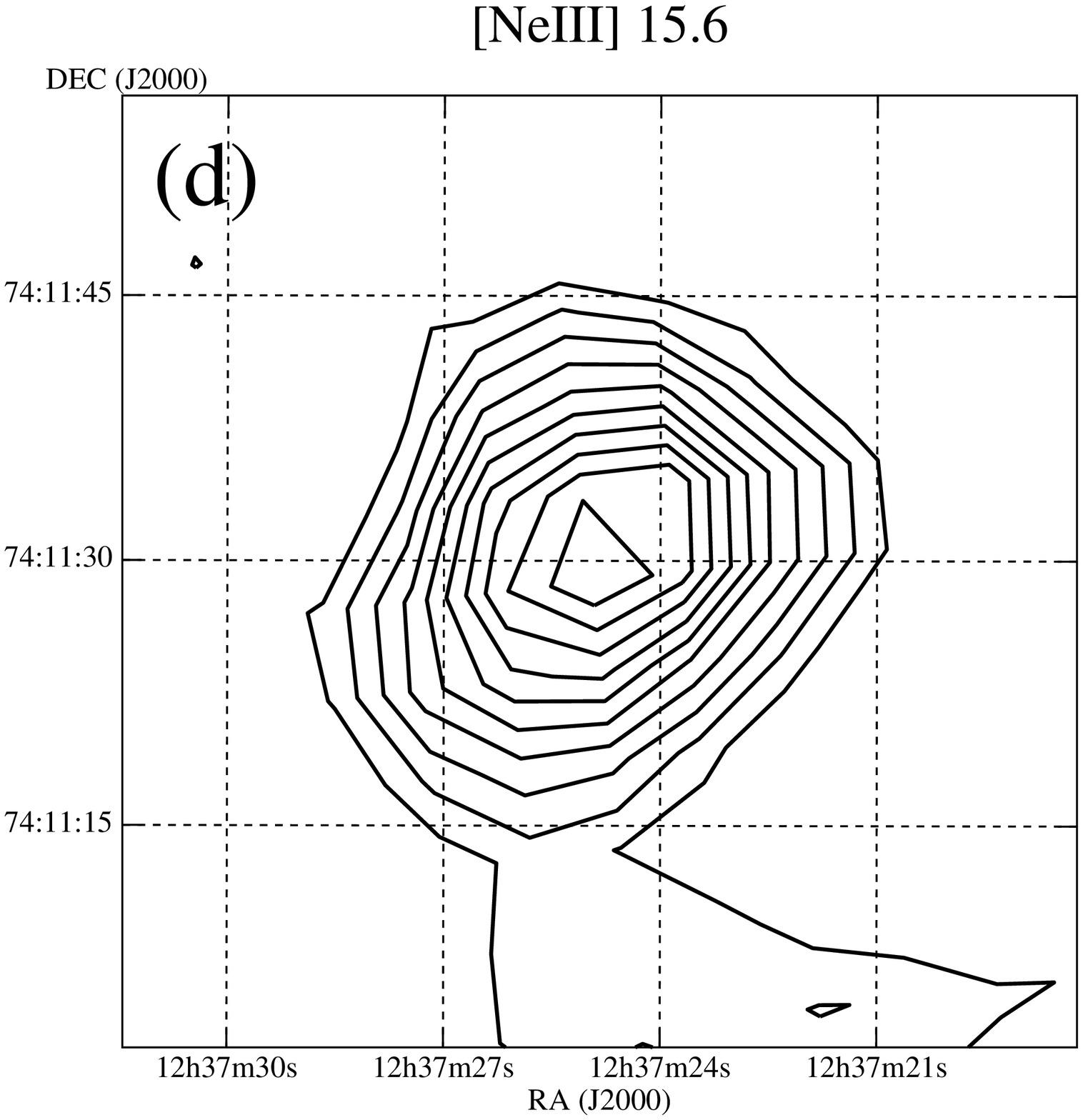}

\plottwo{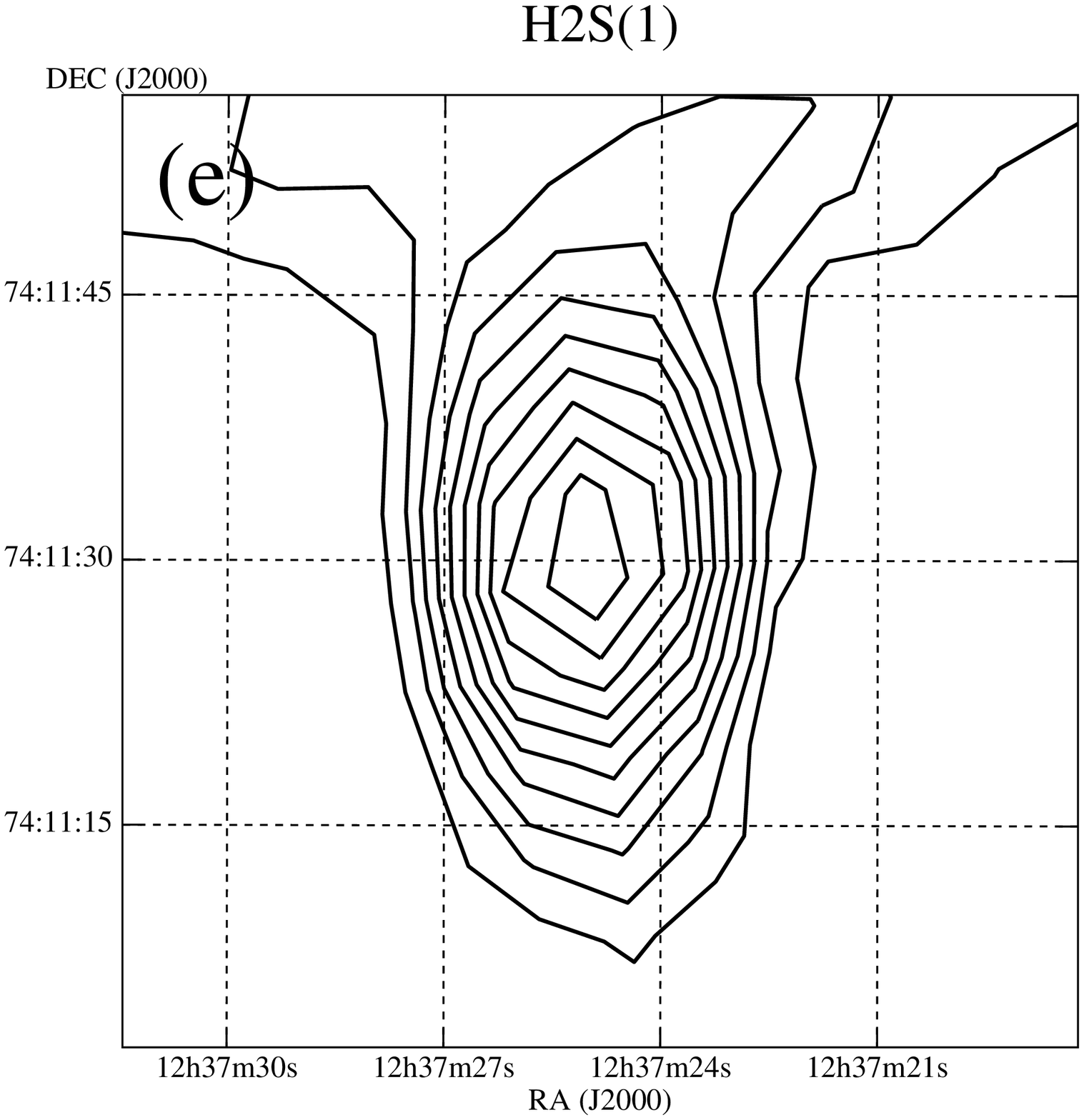}{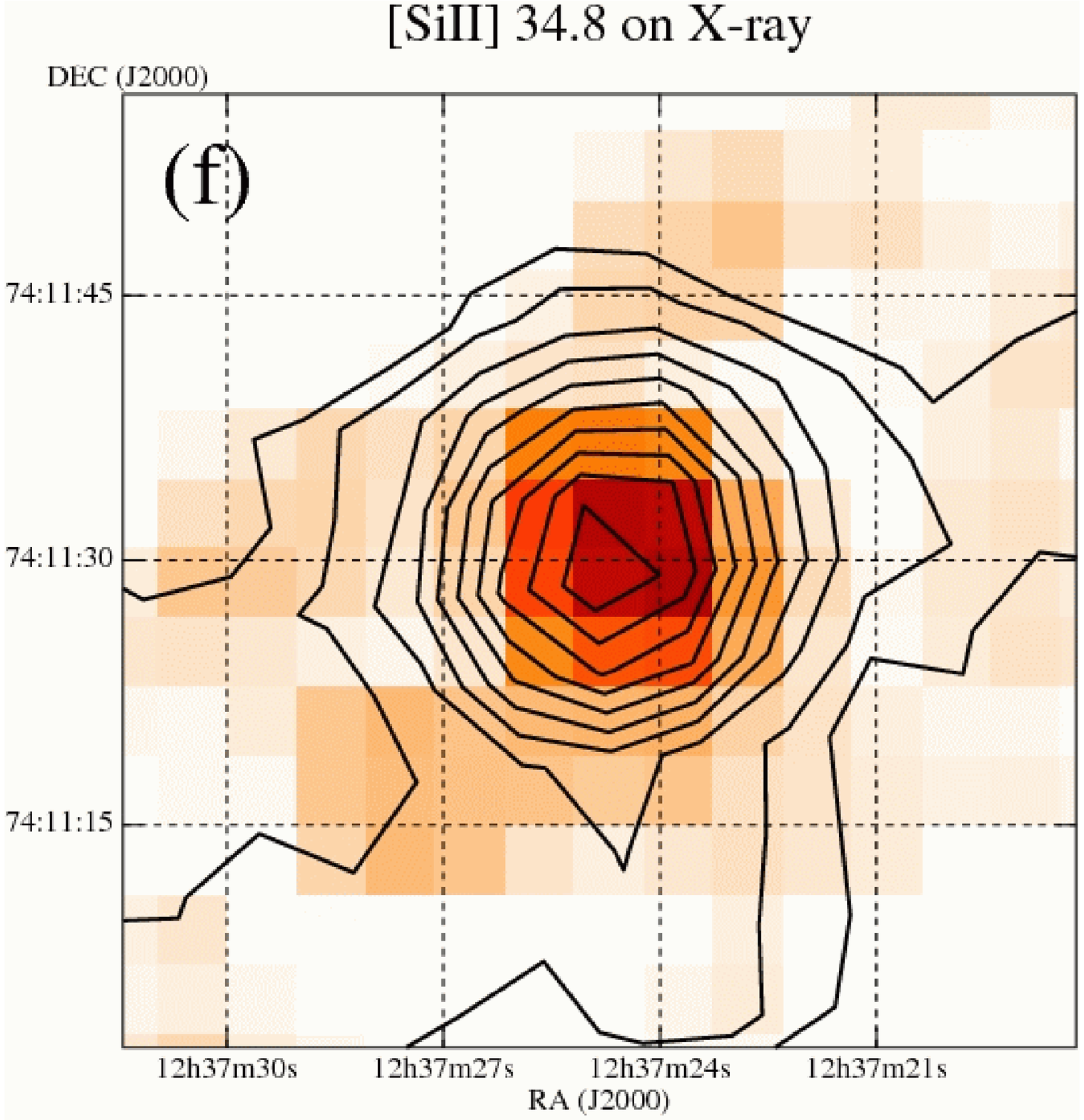}
\caption{Contour maps of (a) the H$_2$S(3), (b) PAH 12.7 $\mu$m, (c) [NeII] 12.8 $\mu$m, (d) [NeIII] 15.6 $\mu$m, (e) H$_2$S(1), and (f) [SiII] 34.8 $\mu$m emissions. The {\it Chandra} X-ray image is superposed on the [SiII] contour map. For each map, the contours are drawn on a linear scale from 10 to 90 \% of the peak (H$_2$S(3): $9.3\times 10^{-9}$ W m$^{-2}$ sr$^{-1}$, PAH 12.7 $\mu$m: $8.7\times 10^{-9}$ W m$^{-2}$ sr$^{-1}$, [NeII]: $5.9\times 10^{-9}$ W m$^{-2}$ sr$^{-1}$, [NeIII]: $1.2\times 10^{-9}$ W m$^{-2}$ sr$^{-1}$, H$_2$S(1): $3.1\times 10^{-9}$ W m$^{-2}$ sr$^{-1}$, [SiII]: $3.4\times 10^{-9}$ W m$^{-2}$ sr$^{-1}$). The lowest contours correspond to levels with S/Ns of 1.7 (H$_2$S(3)), 1.9 (PAH 12.7 $\mu$m), 2.4 ([NeII]), 1.8 ([NeIII]), 2.6 (H$_2$S(1)), and 2.0 ([SiII]).}
\end{figure}

In Fig.3, we show the contour maps of other spectral components, although their S/Ns are rather low as compared to the PAH 11.3 $\mu$m emission in Fig.2. The robust results obtained from these maps are as follows: (1) the molecular hydrogen line emissions come from the dust lane, similarly to the PAH 11.3 $\mu$m feature. (2) The [NeII] line emission shows a more compact distribution near the nucleus than the PAH 12.7 $\mu$m feature. (3) The [SiII] line emission shows a spatial distribution extending to the southeast and the west, similarly to the PAH 17 $\mu$m feature. 

The presence of the H$_2$ lines implies that there is far-UV emission from early-type stars or shocks providing collisional heating. The spatial coincidence with the strong PAH 11.3 $\mu$m feature that reflects soft radiation field supports the latter scenario for heating the molecular gas. Sofue \& Wakamatsu (1993) concluded that the molecular gas is probably situated only in the nucleus, since the velocity dispersion of the CO line, 105 km s$^{-1}$, is significantly lower than the rotation velocity of the dust lane (200 km s$^{-1}$; M\"ollenhoff \& Bender 1989). However we find that the molecular gas is indeed situated in the dust lane. 

In Fig.3f, the X-ray image retrieved from the {\it Chandra} data archive is superposed on the [SiII] contour map, where the hot plasma also appears to extend from the nucleus to the west and the southeast. The early IRS study found a positive correlation between the [SiII] line and the X-ray luminosity from a sample of X-ray-emitting dusty elliptical galaxies \citep{Kan08a}. The spatial correlation in Fig.3f, together with the luminosity correlation, may suggest relatively abundant gas-phase Si in the hot plasma through sputtering destruction of dust grains where Si is depleted. 

The highly-ionized gas responsible for the [NeII] and [NeIII] lines is likely to be associated with a low-luminosity AGN or LINER nucleus \citep{Smi07}, which is consistent with the relatively compact distribution of the [NeII] line emission as compared to that of the PAH 12.7 $\mu$m feature. Alternatively, the highly-ionized gas might be associated with the above hot plasma, since there is a hint of spatial correspondence in the low-level brightness for the [NeIII] line emission. From the difference in spatial distribution between the PAH features and these lines, we can, at least, deduce that the properties of the PAHs are not significantly regulated by hard radiation from the LINER nucleus.
 
In Fig.4, we superpose the contour maps of the PAH 11.3 $\mu$m and 17 $\mu$m features onto the {\it AKARI} 90 $\mu$m image of NGC~4589 \citep{Kan08b}. Though their spatial resolutions are different, the distribution of the far-IR dust emission is extended in similar directions to that of the PAH 17 $\mu$m feature, both of which are apparently different from the stellar distribution. Since NGC~4589 is thought to be a $10^8$-year-old merger remnant, the distributed dust is likely to have been brought in by the merger, and so are the large PAHs emitting the 17 $\mu$m feature. Considering relatively short destruction timescales of the PAHs in the hot plasma, the large PAHs might have be produced later through the fragmentation of large carbonaceous grains \citep{Ona10}. Then, part of the dust and gas are likely to have fallen into the center, which are settled into a ring or a disk (i.e. the dust lane). The distinctive spatial distribution of the PAH 11.3 $\mu$m emission perfectly following the dust lane indicates that the PAHs responsible for the 11.3 $\mu$m feature are not directly brought in by the merger but secondary products created at later evolutionary stages of the brought-in ISM. The similarity in the distribution between the PAH 11.3 $\mu$m feature and the H$_2$ line emission suggests that the PAHs may be created through grain surface chemistry in the same manner as H$_2$ formation in the dense gas region. In the absence of hard radiation field, the PAHs can well exist in superhydrogenated states, which even act as catalysts for interstellar H$_2$ formation \citep{Rau08}.

\begin{figure}
\epsscale{.95}
\plotone{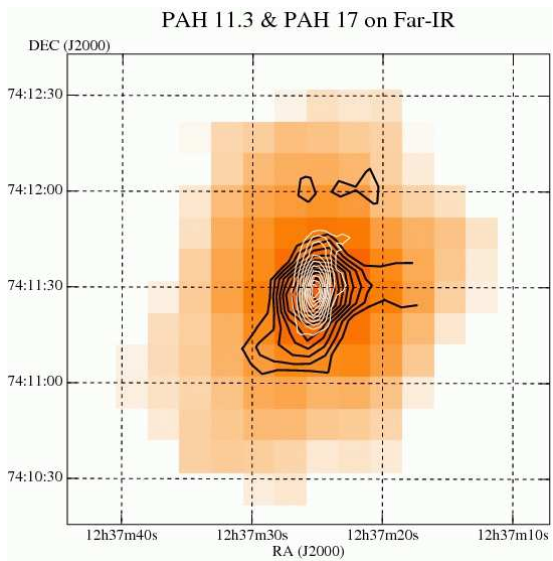}
\caption{AKARI 90 $\mu$m image of NGC~4589 \citep{Kan08b} shown together with the contour maps of the PAH 11.3 $\mu$m (thin white lines) and PAH 17 $\mu$m (thick black lines) features, which are the same as Fig.2.  }
\end{figure}

\section{Conclusions}
With the {\it Spitzer}/IRS, we performed deep spectral mapping observations of NGC~4589, a merger remnant with a minor-axis optical dust lane. As a result, we obtain clear evidence that the PAH 11.3 $\mu$m emission comes predominantly from the optical dust lane of the galaxy. We also detect molecular hydrogen line emissions from the dust lane. The distribution of the PAH 17 $\mu$m emission is extended more widely in directions similar to that of the far-IR dust continuum emission. From their distinctive distributions, we suggest that the PAHs emitting the 11.3 $\mu$m feature are secondary products through the evolution of the ISM brought in by the merger, while the distributed dust and large PAHs emitting the 17 $\mu$m feature are relics of the merger.  

\acknowledgments
This work is based on observations made with the {\it Spitzer} Space Telescope, which is operated by the Jet Propulsion Laboratory, California Institute of Technology under NASA contract 1407. We would like to thank the IRS team and the SSC for their dedicated work in generating the BCD. We are grateful to the SINGS team for providing the CUBISM software. We would also express many thanks to the anonymous referee for giving us useful comments. This work has also made use of data obtained from the {\it Chandra} Data Archive. This research is supported by the Grants-in-Aid for the scientific research Nos. 19740114 and 21740139, and the Nagoya University Global COE Program, ``Quest for Fundamental Principles in the Universe: from Particles to the Solar System and the Cosmos'', both from the Ministry of Education, Culture, Sports, Science and Technology of Japan.


%


\end{document}